\documentclass[prd,aps,twocolumn,superscriptaddress,eqsecnum,floatfix,preprintnumbers,amsmath,amssymb,
nofootinbib,longbibliography]{revtex4-2}

\usepackage{amssymb,amsmath}
\usepackage{epsfig}
\usepackage[dvipsnames]{xcolor}
\usepackage[utf8]{inputenc}
\usepackage{stmaryrd}
\usepackage{mathrsfs}
\usepackage{mathalfa}
\usepackage{accents}
\usepackage[normalem]{ulem}

\usepackage{graphicx}
\graphicspath{ {c:/Documents/PhysicsNotes/GradResearch/Images/GR_Lens/} }

\newcommand{\pa}{\partial}

\newcommand{\be}{\begin{equation}}
\newcommand{\ee}{\end{equation}}
\newcommand{\bea}{\begin{eqnarray}}
\newcommand{\eea}{\end{eqnarray}}
\newcommand{\ba}{\begin{equation}\begin{aligned}}
\newcommand{\ea}{\end{aligned}\end{equation}}

\newcommand{\beg}{\begin{gather*}}
\newcommand{\eng}{\end{gather*}}
\newcommand{\hh}{,\hspace{0.5cm}}
\newcommand{\hhh}{,\hspace{0.2cm}}

\newcommand{\n}[1]{\label{#1}}

\newcommand{\CAL}{\mathcal}

\newcommand{\ts}[1]{{\boldsymbol{#1}}}
\newcommand{\ie}{\emph{i.e.} }

\def\XXint#1#2#3{{\setbox0=\hbox{$#1{#2#3}{\int}$ }
\vcenter{\hbox{$#2#3$ }}\kern-.6\wd0}}

\usepackage[makeroom]{cancel}
\usepackage[caption=false]{subfig}
\usepackage[colorlinks=true,
            citecolor=green,
            linkcolor=red,
            filecolor=cyan,
            urlcolor=magenta,
            backref=false]{hyperref}

\newcommand{\BS}{\begin{split}}
\newcommand{\ES}{\end{split}}

\begin{document}

\title{Motion of  Rotating Black Holes in Homogeneous Scalar Fields: A General Case}

\author{Valeri P. Frolov}%
\email[]{vfrolov@ualberta.ca}
\affiliation{Theoretical Physics Institute, Department of Physics,
University of Alberta,\\
Edmonton, Alberta, T6G 2E1, Canada
}
\author{Alex Koek}
\email[]{koek@ualberta.ca}
\affiliation{Theoretical Physics Institute, Department of Physics,
University of Alberta,\\
Edmonton, Alberta, T6G 2E1, Canada
}


\begin{abstract}
 In this paper we consider the motion of a rotating black hole through a static, homogeneous, massless scalar field. In the general case, a constant vector of the field gradient can be timelike, spacelike or null.
 We consider and compare all of these cases.  We demonstrate that as a result of the interaction of the black hole with the scalar field, its mass, spin and relative velocity with respect to the field can change. We obtain the equations describing the evolution of these parameters and present solutions of the obtained equations for some simple cases.
 \hfill {\scriptsize Alberta Thy 3-24}
\end{abstract}

\maketitle

\section{Introduction}

A scalar field plays an important role in both theoretical physics and cosmology. Scalar fields with spontaneous symmetry breaking are used to explain the origin of mass for particles in the Standard Model.
Quite often in cosmology, a scalar field is used to model the inflaton field which drives inflation. It is also used to describe possible phase transitions in the early Universe. A wide class of interesting models are those which include a shift-invariant scalar field, since these models allow for solutions with a non-vanishing field gradient. For such a state, Lorentz invariance is broken. Examples of such theories are ghost condensate models \cite{Arkani-Hamed:2003pdi,Arkani-Hamed:2003juy}.
A model which contains a constant spacelike gradient of the scalar field was considered in application to cosmology within the framework of a so-called solid inflation model \cite{Endlich:2012pz}.
A scalar field was also used for the description of the phenomenon of emergent time and dynamics in Euclidean space \ (see e.g. \cite{Moffat:1992bf,Babichev:2007dw,Mukohyama:2013ew,Mukohyama:2013gra}).
In the khrono-metric model, a Lorentz invariance-breaking scalar field wass used to label  a foliation of spacetime by a family of spacelike surfaces (see e.g. \cite{Kovachik:2023fos} and references therein). This model is closely related to the Einstein–aether theory \cite{Jacobson_2001,jacobson2008einsteinaether,PhysRevD.97.124023}, in which the preferred frame effects are described by a unit timelike vector $\ts{u}$, called aether.

In this paper, we consider a rotating black hole moving in a scalar field.
When a black holes moves in an external field, the mutual interaction between the black hole and the field changes the state of both components. The configuration of the external field near the black hole changes due to strong gravity in its vicinity. This field can be scattered by the black hole, and can also be partially absorbed. At the same time, the parameters of the moving black hole  change. A moving, rotating, electrically neutral black hole is characterized by its mass $M$ and spin $J$, as well as the velocity $\vec{V}$ of its motion. The main goal of this paper is to obtain equations which determine the evolution of these parameters for a rotating black hole moving in a general-type homogeneous massless scalar field.

In the recent publication \cite{Frolov:2024xyo}, the motion of a rotating black hole in a constant homogeneous electromagnetic field was discussed. Using an exact solution to Maxwell's equations in the Kerr metric for specially adapted asymptotic conditions at flat infinity, the momentum flux of the field into the black hole was calculated. This allows one to obtain the expression for the force acting on the black hole and the torque acting on its spin\footnote{General aspects of the interaction of a rotating black hole with surrounding matter is discussed in the recent paper \cite{Dyson:2024qrq}}. 

The motion of a rotating black hole in a special type of homogeneous scalar massless field was discussed in \cite{Frolov:2023gsl}. It was assumed that a constant gradient of the scalar field is a timelike vector. Surprisingly, it was shown that the behavior of black holes in homogeneous scalar fields is quite different from the behavior of black holes in homogeneous electromagnetic fields. The main difference is the following: A black hole moving in an electromagnetic field retains a constant mass, while a black hole moving in a scalar field grows in mass due to absorption of the field by the black hole. Moreover, the mass of the black hole formally increases to infinity in a finite time interval. One of the goals of this paper is to understand the origin of such a difference between these two cases. 

For this purpose, we consider a general case of the constant homogeneous massless scalar field without imposing the assumption that its gradient is timelike. We demonstrate that the case where the gradient is spacelike shares many common features with the case of the electromagnetic field.  Another new element of this paper is the calculation of all of the components of the torque acting on the spin of the black hole as it moves in a scalar field.  

The paper is organized as follows: After discussion of a static homogeneous massless field in flat spacetime in section~II, we present an exact solution for the scalar field in the presence of a moving black hole in section~III, and calculate the fluxes of the momentum and angular momentum of the field into the black hole. In section~IV, the equations for the evolution of the mass and spin of the black hole are obtained and discussed.
Equations of motion of a rotating black hole moving in the scalar field are derived in section~V. Some interesting solutions of these equations are presented in section~VI. The results obtained in the paper are summarized and discussed in section~VII. In this section we also compare black hole motion in scalar fields versus black hole motion in electromagnetic fields, and briefly discuss possible applications.

In this paper, we use MTW sign conventions \cite{MTW} and units in which $c=G=\hbar=1$.

\section{Homogeneous scalar field in a flat spacetime}

Let $\Psi$ be a  massless scalar field obeying  the following equation
\be 
\Box \Psi =0\, ,
\ee
in flat spacetime. Its stress-energy tensor is 
\be\n{SET}
{T}_{\mu\nu}={\Psi}_{,\mu}{\Psi}_{,\nu} - \frac{1}{2}g_{\mu\nu}{\Psi}_{,\alpha}{\Psi}^{,\alpha}\,.
\ee

We consider a special class of so-called homogeneous solutions for which the gradient of $\Psi$ is a constant vector. There exist 3 different types of these solutions, which we denote as $T$-, $S$-, and $N$-fields. For $T$-fields the gradient vector is timelike, for $S$-fields this vector is spacelike, and for $N$-fields it is null.

Using Lorentz transformations, it is possible to find a reference frame $\tilde{K}$ in which these solutions have the following  form
\be \n{KK}
\nabla_{\mu}\Psi=\Psi_0 
\begin{cases}
(1,0,0,0)&  \text{for $T$-field} ,   \\
(0,N_X,N_Y,N_Z)&  \text{for $S$-field} ,   \\
(1,N_X,N_Y,N_Z)&  \text{for $N$-field}   ,
\end{cases}
\ee
where $\vec{N}=(N_X,N_Y,N_Z)$ is a 3D unit vector.
It is convenient to keep the direction of the vector $\vec{N}$ not fixed, and to use this ambiguity later to simplify relations which appear. We denote by $\tilde{K}$ a frame in which the gradient of the field has the canonical form \eqref{KK} and call it the ``field frame". We also denote by $\tilde{X}^{\mu}=(\tilde{T},\tilde{X},\tilde{Y},\tilde{Z})$ Cartesian coordinates in this frame.
One has
\be 
\nabla_{\mu}\Psi \nabla^{\mu}\Psi=\epsilon\Psi_0^2 \, ,
\ee
where $\epsilon$ takes values $-1$, $+1$ and $0$ for $T$-, $S$-, and $N$-solutions, respectively.

Certainly, it is possible by a rigid rotation of the coordinate axes to make the $x$-axis to be directed along $\vec{N}$. For this choice of coordinates, one has
\be \n{SKK}
\Psi= 
\begin{cases}
\Psi_0 \tilde{T}&  \text{for $T$-field} ,   \\
\Psi_0 \tilde{X}&  \text{for $S$-field} ,   \\
\Psi_0 (\tilde{T}+\tilde{X})&  \text{for $N$-field}   .
\end{cases}
\ee
The stress-energy tensor for the $T$-, $S$-, and $N$-fields in the $\tilde{K}$ frame has the form
\be \n{TKX}
\tilde{T}_{\mu\nu}=\Psi_0^2 
\begin{cases}
\frac{1}{2}\,\text{diag}(1,1,1,1)&  \text{for $T$-field} ,   \\
\frac{1}{2}\,\text{diag}(1,1,-1,-1)&  \text{for $S$-field} ,   \\
\delta_{\mu}^{\tilde{T}} \delta_{\nu}^{\tilde{T}}+
\delta_{\mu}^{\tilde{X}} \delta_{\nu}^{\tilde{X}}+
2\delta_{(\mu}^{\tilde{T}} \delta_{\nu)}^{\tilde{X}}&  \text{for $N$-field}  \, .
\end{cases}
\ee

Before we proceed further, let us make some comments about special features of the $N$-field solution. Its distinguishing property is that a canonical frame $\tilde{K}$ is not uniquely defined by \eqref{KK}. To show this let us denote
\be 
\tilde{u}=\dfrac{1}{\sqrt{2}}(\tilde{T}-\tilde{X})\hh 
\tilde{v}=\dfrac{1}{\sqrt{2}}(\tilde{T}+\tilde{X})\, .
\ee 
The stress-energy tensor for such a field is
\be 
\tilde{T}_{\mu\nu}=2\Psi_0^2 \tilde{v}_{, \mu}\tilde{v}_{, \nu}\,.
\ee
This tensor describes a null fluid with energy density $\sim \Psi_0^2$ propagating in the negative $\tilde{X}$-direction.
Under a Lorentz boost along the positive $\tilde{X}$-axis, the retarded
and advanced null coordinates $\tilde{u}$ and $\tilde{v}$ transform as follows
\be 
v=\alpha^{-1} \tilde{v}\hhh u=\alpha \tilde{u}\hhh \alpha=\sqrt{\dfrac{1+V}{1-V}}\, .
\ee
Here $V$ is the boost velocity. Hence, the Lorentz boost transformation preserves the form of the $N$-field with the only change being $\Psi_0\to\alpha\Psi_0$.
The stress-energy tensor after this boost takes the form
\be \n{TRED}
T_{\mu\nu}=2 \alpha^{2} \Psi_0^2 {v}_{, \mu}{v}_{, \nu}\, .
\ee 

Let us make another remark concerning the definition of the $N$-field solution. In \eqref{SKK} we assume that the vector $\vec{N}$ is along the positive $\tilde{X}$-axis. For the choice of $\vec{N}$ in the opposite direction, the form of the solution slightly changes. To distinguish these two solutions, we use the following notation
\be 
\Psi_N^{\pm}=\Psi_0(\tilde{T}\pm \tilde{X})\, .
\ee
For the $\Psi_N^-$ field its ``null fluid" flux is ``right-moving" (\ie it is propagating in the positive $\tilde{X}$-direction). Let us note that $T$-field can be presented as a superposition of $\Psi_N^{\pm}$ fields
\be 
\Psi_T=\dfrac{1}{2}(\Psi_N^+ +\Psi_N^-)\, ,
\ee 
and for this state, the positive and negative energy fluxes are compensated so that the off-diagonal components of the stress-energy tensor vanish.

After these remarks, we  continue by using the coordinates $\tilde{X}^{\mu}$ in which  the gradient of the field $\Psi$ has the form \eqref{KK}.
Consider another frame $K$ which moves with respect to the field frame $\tilde{K}$ with velocity $\vec{V}$. We choose its coordinate axes to be parallel to the axes of $\tilde{K}$, and denote by $X^{\mu}=(T,X,Y,Z)$ the Cartesian coordinates in this frame.

The coordinates $X^{\mu}$ and $\tilde{X}^{\mu}$ are related as follows
\be \n{KKK}
\begin{split}
& \tilde{T} = \gamma \big(  
T+\vec{V}\cdot \vec{r}\big)\, ,\\
& \tilde{\vec{r}} = \vec{r} +(\gamma-1)\dfrac{\vec{V}\cdot \vec{r}}{V^2}\vec{V} +\gamma T \vec{V}\, .
\end{split}
\ee
Here, $\vec{r}=(X,Y,Z)$, $\tilde{\vec{r}}=(\tilde{X},\tilde{Y},\tilde{Z})$, and $\gamma$ is the Lorentz factor,
\be 
\gamma=\dfrac{1}{\sqrt{1-V^2}}\, .
\ee 
The homogeneous scalar field solution in $K$ frame is of the form
\be \n{PPKK}
\Psi=\Psi_0 (U_{T} T+U_X X +U_Y Y +U_Z Z)\, .
\ee 
In what follows, we identify $K$ with the asymptotic frame in which the black hole is at rest, and call it the ``BH-frame".

Using \eqref{KKK}, it is easy to find the expressions for the parameters $\{ U_{T}, U_X, U_Y, U_Z\}$ in terms of the velocity vector $\vec{V}$.  For each field solution, we have the following.
\begin{itemize}
\item $T$-field
\be 
U_{T}=\gamma\hhh \vec{U}=\gamma \vec{V}\, .
\ee
\item $S$-field
\be
\begin{split}
&U_{T}=\gamma (\vec{N}\cdot\vec{V})\, ,\\ 
& \vec{U}=\vec{N}+(\gamma-1)(\vec{N}\cdot\vec{V})\dfrac{\vec{V}}{V^2}\, .
\end{split}
\ee
\item $N$-field
\be
\begin{split}
& U_{T}=\gamma(1+\vec{N}\cdot\vec{V})\,,\\
&\vec{U}=\vec{N}+\gamma\vec{V}+(\gamma-1)(\vec{N}\cdot\vec{V})\dfrac{\vec{V}}{V^2}\, .
\end{split}
\ee
\end{itemize}

\section{Energy-momentum and angular momentum fluxes in the $BH$-frame}

\subsection{Kerr metric}

The Kerr metric, describing a vacuum stationary rotating black hole, written in Boyer-Lindquist coordinates, is
\begin{equation}\label{Kerr}
\begin{split}
ds^2 &= -\left( 1-\frac{2Mr}{\Sigma}\right) dt^2
-\frac{4Mar\sin^2\theta}{\Sigma} dt d\varphi\\
&+\dfrac{A\sin^2\theta}{\Sigma} d\varphi^2+\frac{\Sigma}{\Delta} dr^2 +\Sigma d\theta^2\, ,\\
\Sigma&=r^2+a^2\cos^2\theta\hh \Delta=r^2-2Mr+a^2 \, ,\\
A&=(r^2+a^2)^2-a^2\Delta \sin^2\theta\\
&=\Sigma(r^2+a^2)+2M a^2 r\sin^2\theta\, .
\end{split}
\end{equation}
Here $M$ is the mass of the black hole, and $a$ is its rotation parameter related to the black hole's spin by $J=Ma$.
The metric has two commuting Killing vectors $\ts{\xi}=\pa_t$ and $\ts{\zeta}=\pa_{\varphi}$. Let us denote
\be
r_{\pm}=M\pm \sqrt{M^2-a^2}\, .
\ee
The black hole's outer horizon is located at $r=r_+$.

For $M=0$, the curvature vanishes and the metric \eqref{Kerr} takes the form
\be
ds^2=-dt^2+\dfrac{\Sigma}{r^2+a^2}dr^2+\Sigma d\theta^2 +(r^2+a^2)\sin^2\theta d\varphi^2\, .
\ee
This is nothing but a flat metric in oblate spheroidal coordinates $(t,r,\theta,\varphi)$, which are related to the flat Cartesian coordinates $(T,X,Y,Z)$ as follows
\be
\begin{split}
&T=t\, ,\\
&X=\sqrt{r^2+a^2}\sin\theta\cos\varphi\, ,\\
&Y=\sqrt{r^2+a^2}\sin\theta\sin\varphi\, ,\\
&Z=r\cos\theta\, .
\end{split}
\ee

The Killing vectors generating translations along the $T$, $X$, $Y$, and $Z$ coordinate axes written in oblate spheroidal coordinates $(t,r,\theta,\phi)$  are
\ba\n{TRA}
\xi^{\mu}_{(T)}\partial_{\mu} =& \partial_{t}\,,\\
\xi^{\mu}_{(X)}\partial_{\mu} =& \frac{\sqrt{r^2+a^2}}{\Sigma}\cos{\varphi}\Big(r\sin{\theta}\,\partial_{r}+\cos{\theta}\,\partial_{\theta}\Big)\\
&-\frac{\sin{\varphi}}{\sqrt{r^2+a^2}\sin{\theta}}\,\partial_{\varphi}\,,\\[5pt]
\xi^{\mu}_{(Y)}\partial_{\mu} =& \frac{\sqrt{r^2+a^2}}
{\Sigma}\sin{\varphi}\Big(r\sin{\theta}\,\partial_{r}+\cos{\theta}\,\partial_{\theta}\Big)\\
&+\frac{\cos{\varphi}}{\sqrt{r^2+a^2}\sin{\theta}}\,\partial_{\varphi}\,,\\[5pt]
\xi^{\mu}_{(Z)}\partial_{\mu} =& \frac{(r^2+a^2)\cos{\theta}}{\Sigma}\,\partial_{r}-\frac{r\sin{\theta}}{\Sigma}\,\partial_{\theta}\,.
\ea

The Killing vectors generating rotations about the $X$-, $Y$- and $Z$-axes written in oblate spheroidal coordinates are
\be\n{ROT}
\begin{split}
\zeta_{(X)}^{\mu}\pa_{\mu}&=\dfrac{\sqrt{r^2+a^2}}{\Sigma}\sin\varphi
\big[ a^2\sin\theta\cos\theta\ \pa_r-r\pa_{\theta}\big]\\
 & -\dfrac{r\cos\theta\cos\varphi}{\sqrt{r^2+a^2}\sin\theta}\ \pa_{\varphi}\, ,\\
\zeta_{(Y)}^{\mu}\pa_{\mu}&=-\dfrac{\sqrt{r^2+a^2}}{\Sigma}\cos\varphi
\big[a^2 \sin\theta\cos\theta\ \pa_r- r\pa_{\theta} \big] \\
& -\dfrac{r\cos\theta \sin\varphi }{\sqrt{r^2+a^2}\sin\theta}\ \pa_{\varphi}\, ,\\
\zeta_{(Z)}^{\mu}\pa_{\mu}&=\pa_{\varphi}\, .
\end{split}
\ee

The Boyer-Lindquist coordinates are singular at the black hole horizon. To cover the future horizon and the interior of the black hole, one can use the incoming Kerr coordinates\footnote{
Since the radial coordinate $r$ has dimensions of length, formally the quantities similar to $\ln r$ should have the form $\ln(r/\ell)$, where $\ell$ is a constant length-scale parameter. For example, one can choose $\ell=M$. Let us notice that the change of the scale $\ell$ can always be absorbed by a constant shift of the parameters $t$, $v$, $f$, and terms similar to them. Since the concrete choice of $\ell$ is not important for further results, we simply choose to omit it.
}
\be\n{KIN}
\begin{split}
&v=t+f(r)\, ,\\
&f(r)=\int \dfrac{(r^2+a^2) dr}{\Delta}\\
&= r+\dfrac{2M}{r_+-r_-} [r_+\ln(r-r_+)-r_-\ln(r-r_-)]\, ,\\
&\psi=\varphi+\varphi_0(r)\, ,\\
&\varphi_0(r)=a\int \dfrac{ dr}{\Delta}=\frac{a}{2\sqrt{M^2-a^2}}\ln\big(\frac{r-r_{+}}{r-r_{-}}\big).
\end{split}
\ee

\subsection{Scalar field in the $BH$-frame}

We start with the following ansatz for the scalar field in Boyer-Lindquist coordinates $(t,r,\theta,\varphi)$:

\ba\n{PPP}
&\Psi = {\Psi}_{0} \Big(U_{T}\mathcal{U}_{T}+U_{X}\mathcal{U}_{X}+U_{Y}\mathcal{U}_{Y}+U_{Z}\mathcal{U}_{Z}\Big)\, ,\\
&\mathcal{U}_{T} = t-r+f(r)-2M\ln(r-r_-)\\
&=v-r-2M\ln(r-r_-)\, ,\\
&\mathcal{U}_{X} = \sin{\theta}\,\big((r-M)\cos\psi-a\sin\psi\big)\,,\\
&\mathcal{U}_{Y} = \sin{\theta}\,\big((r-M)\sin\psi+a\cos\psi\big)\,,\\[4pt]
&\mathcal{U}_{Z} = (r-M)\cos(\theta)\, .
\ea
Here $v$ and $\psi$ are the incoming Kerr coordinates defined in \eqref{KIN}.

The functions $\mathcal{U}_{T}$, $\mathcal{U}_{X}$, $\mathcal{U}_{Y}$, and $\mathcal{U}_{Z}$ which enter the expression \eqref{PPP} for the field $\Psi$ have the following properties
\begin{enumerate}
\item These functions are exact solutions of the massless field equation in the Kerr metric;
\item They are regular at the future event horizon;
\item At large $r$ they have the following asymptotics
\be 
\mathcal{U}_{T}\approx T\hhh 
\mathcal{U}_{X}\approx X\hhh
\mathcal{U}_{Y}\approx Y\hhh
\mathcal{U}_{Z}\approx Z\hhh
\ee
\end{enumerate}
These properties imply that the function $\Psi$ defined by \eqref{PPP} is nothing but a deformation of the homogeneous scalar field \eqref{PPKK} due to the presence of a rotating black hole.
The regularity of the functions $\mathcal{U}_{T}$, $\mathcal{U}_{X}$, $\mathcal{U}_{Y}$, and $\mathcal{U}_{Z}$  on the horizon can be easily checked by writing the  expressions for these quantities in the incoming Kerr coordinates \eqref{KIN}.

\subsection{Fluxes}

Let $\eta^{\mu}$ be a vector in the Kerr spacetime in Boyer-Lindquist coordinates $x^{\mu}=(t,r,\theta,\varphi)$, and let $T_{\mu\nu}$ be the stress-energy tensor of the field. 
We define a current
\be 
K_{\mu}=\eta^{\nu}T_{\mu\nu}\, .
\ee
Let $S_0$ be a 2D surface in the Kerr spacetime defined by the equations  $t=\mbox{const.}$ and $r=r_0=\mbox{const.}$, and let $\Sigma$ be a 3D surface obtained by the shift of $S_0$ 
during the time interval $t\in(t_1,t_2)$. 
The flux of the current $K_{\mu}$ through $\Sigma$ is
\be \n{KI}
K=\int K_{\mu} d\Sigma^{\mu}\, ,
\ee
where 
\be 
d\Sigma_{\mu}=\dfrac{1}{6} \sqrt{-g}\, \epsilon_{\mu\beta_1,\beta_2\beta_3}\mbox{det}\Bigg(
\dfrac{\pa x^{\beta_i}}{\pa y_j}\Bigg) dt\, d\theta d\varphi
\, .
\ee 

Here, $\sqrt{-g}=\Sigma \sin\theta$, $\epsilon_{\mu\beta_1\beta_2\beta_3}$ is a Levi-Civita symbol with $\epsilon_{0 1 2 3}=1$, and $y^j=(t,\theta,\varphi)$ are coordinates on $\Sigma$. For this choice of coordinates, one has
\be 
d\Sigma^{\mu}=-\Delta \delta^{\mu}_{r} dt\,d\omega\hh 
d\omega=\sin\theta d\theta\, d\varphi\, .
\ee 
The sign in the definition of the volume element $d\Sigma^{\mu}$ is chosen so that \eqref{KI} determines the flux through $\Sigma$ into its interior.

We assume that the current $K^{\mu}$ is stationary, $\CAL{L}_{\xi}K_{\mu}=0$, and so one has
\be\n{TKK} 
K=(t_2-t_1) \CAL{K}\hh 
\CAL{K}=-\Delta \int_{S_0} \eta^{\mu} T_{\mu r} d\omega\, .
\ee
Here $\CAL{K}$ is the flux $K$ calculated per a unit interval of time.

If $\ts{\eta}$ is a Killing vector, then $\CAL{K}$ does not depend on $r_0$, and more generally, it does not depend on the choice of the 2D surface $S$ surrounding the black hole. In particular, these fluxes can be calculated directly on the horizon of the black hole. 

\subsubsection{Energy flux}

Let us first discuss the energy flux into the black hole. We note that in flat spacetime, the flux of the energy in the direction of $i$-th axis is given by the $T^{0i}$ component of the stress-energy tensor. This quantity can be written as $-\xi^{\mu}T_{\mu i}$,
where $\xi^{\mu}$ is a future-directed Killing vector generating time translations. Using this sign convention, we obtain the following expression for the energy flux into the interior of the surface $S_0$
\be 
\CAL{E}=\Delta\int_{r=r_0}  T_{t r} d\omega\, .\\
\ee
For the field \eqref{PPP}, only the $\CAL{U}_T$ component contributes. We have 
\be 
T_{t r}=\pa_t\Psi \pa_r\Psi\,.
\ee 
Only $\CAL{U}_T$ contains dependence on $t$. On the other hand,
all other components $\CAL{U}_i$ change  signs under reflections  $\theta\to\pi-\theta$  and  $\varphi\to\pi+\varphi$. As a result, the contribution from these components after integration over $d\omega$ vanishes.

A simple calculation gives
\be 
\pa_t \CAL{U}_T=1\hh 
\pa_r \CAL{U}_T=\dfrac{2Mr_+}{\Delta}.
\ee
Using these relations, one gets
\be 
\mathcal{E} = \,8\pi{\Psi}_{0}^{2}Mr_{+}U_{T}^{2}\, .
\ee 
The energy flux does not depend on $r_0$, as it should be.

\subsubsection{Momentum and angular momentum fluxes.}

We use \eqref{TKK} to define the energy-momentum and angular momentum fluxes. For this purpose, we use vectors  $\xi^{\mu}_{(\nu)}$
and $\zeta^{\mu}_{(\nu)}$ defined by the relations \eqref{TRA} and \eqref{ROT}. Since these vectors generate asymptotic symmetries, in the definition of the corresponding fluxes we choose the radius $r_0$ of the 2D surface $S_0$ to be large, and take the limit $r_0\to\infty$. Thus, we have
\be \n{EPJ}
\begin{split}
&\CAL{P}_{i}=-\lim_{r_0\to\infty}\Big(\Delta\int_{r=r_0}  \xi_{(i)}^{\mu} T_{\mu r} d\omega
\Big)\, ,\\
& \CAL{J}_{i}=-\lim_{r_0\to\infty}\Big(\Delta\int_{r=r_0}  \zeta_{(i)}^{\mu} T_{\mu r} d\omega
\Big)\, .
\end{split}
\ee 
Let us note that  $\ts{\zeta}\equiv \ts{\zeta}_{(Z)}$ is a Killing vector, and hence $\CAL{J}_Z$ does not depend on $r_0$.

Using \eqref{PPP} and the definition of the stress-energy tensor \eqref{SET}, one finds the expressions which enter as integrands in \eqref{EPJ}. For calculations we use the GRTensor program. In fact, for the calculation of  $\CAL{P}_i$, it is sufficient to expand $T_{\mu\nu}$ in powers of $1/r$ and keep only the terms up to the second order of this expansion. For the calculation of $\CAL{J}_i$, one can proceed in the same way, but terms of the order $1/r^3$ should be included as well.

Performing these rather long but straightforward calculations, one obtains the following expressions for the fluxes of the momentum 
\ba\n{EMF}
\mathcal{P}_{X} = &-8\pi{\Psi}_{0}^{2}MU_{T}\big(r_{+}U_{X} + \frac{1}{3}a U_{Y}\big)\,,\\[8pt]
\mathcal{P}_{Y} = &-8\pi{\Psi}_{0}^{2}MU_{T}\big(r_{+}U_{Y} - \frac{1}{3}a U_{X}\big)\,,\\[8pt]
\mathcal{P}_{Z} = &-8\pi{\Psi}_{0}^{2}Mr_{+}U_{T}U_{Z}\,.\\[8pt]
\ea

Similarly, for the fluxes of the angular momentum, one finds
\ba\n{AMF}
\mathcal{J}_{X} = &\,\frac{4}{3}\pi{\Psi}_{0}^{2}Ma U_{Z}\Big(MU_{X} + \frac{8}{5}a U_{Y} \Big)\,,\\[8pt]
\mathcal{J}_{Y} = &\,\frac{4}{3}\pi{\Psi}_{0}^{2}Ma U_{Z}\Big(MU_{Y} - \frac{8}{5}a U_{X} \Big)\,,\\[8pt]
\mathcal{J}_{Z} = &-\frac{4}{3}\pi{\Psi}_{0}^{2}M^{2}a\big(U_{X}^{2}+U_{Y}^{2}\big)\,.
\ea

\section{Change of the black hole parameters}

\subsection{Force and torque in the BH-frame}

The flux of the energy-momentum changes the energy and momentum of the black hole. The corresponding 4D vector $\CAL{F}^{\mu}$ of the force in the $BH$-frame is 
\be 
\CAL{F}^{\mu} = (\mathcal{E},\mathcal{P}_{X},\mathcal{P}_{Y},\mathcal{P}_{Z})\, .
\ee 
Similarly, the torque acting on the spin of the black hole is
\be 
\vec{\CAL{T}}=(\mathcal{J}_X,\mathcal{J}_Y,\mathcal{J}_Z)\, .
\ee

Let us recall that we did not specify the direction $\vec{N}$ of the gradient of the field. In the reference frame $K$ there exists a preferred direction to orient the frame, namely the axis of the rotation of the black hole. We choose the $Z$-axis of our coordinates to coincide with this direction. For this choice, the spin vector $\vec{J}$ of the black hole is
\be 
\vec{J}=J\vec{n}\hh J=Ma\hhh \vec{n}=(0,0,1)\, .
\ee

Let us denote
\be 
\beta=8\pi\Psi_0^2 \, .
\ee 
Using the expressions for the energy-momentum fluxes in $K$ frame, one can write 4D vector of force acting on the black hole in the following form
\ba
&\CAL{F}^{0} = \beta Mr_+ U_{T}^2\,,\\
&\vec{\CAL{F}} = -\beta U_{T}\Big(Mr_+\vec{U}-\frac{1}{3}\, \vec{J}\times\vec{U}\Big)\,,
\ea
A similar vector form of the expression for the torque is
\be\n{TOR}
\vec{\mathcal{T}} = \frac{1}{6}\beta \bigg[
M \vec{U}\times(\vec{U}\times\vec{J})
+\dfrac{8}{5M} (\vec{U}\cdot\vec{J})
(\vec{U}\times\vec{J})\bigg]\,
\ee

Let us give explicit expressions of the force for the $T$-, $S$- and $N$-fields:
\begin{itemize}
\item $T$-field case
\be \n{FFT}
\begin{split}
&\CAL{F}^{0}_{T}=\beta \gamma^2 Mr_{+}\, ,\\
&\vec{\CAL{F}}_{T}=-\beta\gamma^2\Big[ 
Mr_{+}\vec{V}-\dfrac{1}{3} (\vec{J}\times\vec{V})
\Big]\, .
\end{split}
\ee 
\item $S$-field case
\be \n{FFS}
\begin{split}
&\CAL{F}^{0}_{S}=\beta\gamma^2 Mr_{+}(\vec{N}\cdot\vec{V})^2\, ,\\
&\vec{\CAL{F}}_{S}=-\beta\gamma (\vec{N}\cdot\vec{V})
\Bigg[ 
Mr_{+}\Big(\vec{N}+(\gamma-1)(\vec{N}\cdot\vec{V})\dfrac{\vec{V}}{V^2}\Big)\\ 
&-\dfrac{1}{3} 
\Big(
(\vec{J}\times\vec{N})+(\gamma-1)(\vec{N}\cdot\vec{V})\dfrac{\vec{J}\times\vec{V}}{V^2}
\Big)
\Bigg]\, .
\end{split}
\ee 
\item $N$-field case
\be \n{FFN}
\CAL{F}^{\mu}_{N}=(1+\vec{N}\cdot\vec{V})\Big(\CAL{F}^{\mu}_{T}+\frac{1}{\vec{N}\cdot\vec{V}}\CAL{F}^{\mu}_{S}\Big)\,.
\ee
\end{itemize}

\subsection{Change of the mass and spin of the black hole}

In the $K$ frame in which the black hole is (instantly)  at rest, one has
\be\n{MJ}
\dfrac{dM}{d\tau}=\CAL{F}^{0} \hh 
\frac{d\vec{J}}{d\tau} = \vec{\mathcal{T}}\,.
\ee
These equations describe the change of the mass and spin of the black hole due to its interaction with the scalar field. 
Since $\CAL{F}^{0}= \beta Mr_{+} U_{T}^2\ge 0$, the mass of the black hole $M$ never decreases. It remains constant only when $U_{T}=0$.

In the general case, the value of the spin $J$, as well as its direction $\vec{n}$, changes. To find the rate of change of $J$, we use the following relations
\be 
J\dfrac{dJ}{d\tau}=\dfrac{1}{2}\dfrac{dJ^2}{d\tau}
=\dfrac{1}{2}\dfrac{d\vec{J}^2}{d\tau}=\vec{J}\cdot  \dfrac{d\vec{J}}{d\tau}\, .
\ee
Using relations \eqref{MJ} and \eqref{TOR} one gets
\be \n{DDJJ}
\dfrac{dJ}{d\tau}=-\frac{1}{6}\beta MJ \, [\vec{U}^2-(\vec{U}\cdot \vec{n})^2]\, .
\ee
This relation  shows that the spin value $J$ either decreases or remains the same. The latter case occurs when the black hole's  spin direction vector $\vec{n}$ is parallel to $\vec{U}$. 

Let us emphasize that at this point we assume that the relative rate of change
of the black hole's characteristics is small, \ie $\dot{M}\ll 1$ and $\dot{J}/J\ll 1/M$. This means that during the characteristic black-hole ``relaxation" time $\sim M$, the change of $M$ and $J$ is small and one can describe a black hole geometry by using the Kerr metric with slowly changing mass and spin.

Let us specify the obtained results for the change of the black hole parameters for $T$-, $S$-, and $N$-fields
\begin{itemize}
\item For the $T$-field, the mass of the black hole always grows, while the spin can remains unchanged provided it is parallel to the velocity $\vec{V}$ of the black hole.
\item For the $S$-field, both mass and spin of the black hole can remain the same. This happens when the velocity $\vec{V}$ is orthogonal to the direction $\vec{N}$ of the gradient of the scalar field, and the spin is parallel to $\vec{N}$.
\item For the $N$-field, the mass of the black hole always grows, while the spin can remains unchanged provided the following condition is satisfied
\be \n{Ncondition}
\vec{U}\times\vec{n}=0\, .
\ee    
\end{itemize}

\subsection{Black hole entropy and the second law}

As we showed in the previous section, as the black hole moves through the scalar field its mass either increases or, under special circumstances, remains the same. Similarly, the value of the black hole spin cannot increase. Hence, the radius of the horizon $r_+$ does not decrease. This means that a similar conclusion is valid for the surface area of the black hole $\CAL{A}=8\pi Mr_+$. It remains the same only if both quantities, the mass $M$ and spin $J$, remain the same. 

The entropy of the black hole is proportional to its area. Thus, the black hole entropy does not decrease as the rotating black hole moves in the homogeneous scalar field, as it should be. Let us obtain a relation describing the change of the black hole's entropy in this process.

The second law of black hole physics implies that\footnote{In this relation $S$ is dimensionless entropy obtained from the the ``standard" thermodynamical entropy via division the latter by the Boltzmann constant $k_B$. Similarly, the temperature is defined in ``energy" units and it is obtained by the multiplication of the ``standard" thermodynamical temperature by $k_B$. }
\be 
TdS=dM-\Omega dJ\, .
\ee
Here, $S=\dfrac{1}{4}\CAL{A}$ is the  entropy of the black hole, $\Omega=a/2Mr_{+}$ is its angular velocity, and
\be 
T=\dfrac{2\sqrt{M^2-a^2}}{\CAL{A}}
\ee
is its temperature. Substituting into the relation
\be 
T\dfrac{dS}{d\tau}=\dfrac{dM}{d\tau}-\Omega \dfrac{dJ}{d\tau}
\ee
expression \eqref{MJ} for $dM/d\tau$, and expression \eqref{DDJJ} for 
$dJ/d\tau$, one gets
\be 
T\dfrac{dS}{d\tau}=\frac{1}{2}\beta\Bigg[
2Mr_+ U_{T}^2+\dfrac{1}{6}\dfrac{Ma^2}{r_+}\Big(\vec{U}^2-(\vec{U}\cdot \vec{n})^2 \Big)\Bigg]\, .
\ee
Both terms in the square brackets are non-negative. Hence, the entropy of the rotating black hole moving in the homogeneous scalar field either grows or remains the same. The latter case is only possible when $U_{T}=0$, and the spin  of the black hole $J\vec{n}$ is parallel to the vector $\vec{U}$.

\section{Force acting on the black hole in the field frame}

For an observer located at far distance $L$ from the black hole, it can be described as a small-size rotating object with mass $M$ and spin $\vec{J}$, moving with velocity $\vec{V}$ with respect to this frame. Its equation of motion is of the form
\be
\dfrac{d P^{\mu}}{d\tau}=f^{\mu}\, .
\ee 
Here $P^{\mu}=(M\gamma,M\gamma\vec{V})$ is the four momentum of the black hole in the field frame, and $f^{\mu}$ is the 4D vector of the force in the field frame. 

We denote by 
$\ts{\CAL{F}}=(\CAL{F}^{0},\vec{\CAL{F}})$ the 4D force vector acting on the black hole in the instantly comoving frame $K$.
To obtain the expression for the 4D force $f^{\mu}$ acting on the black hole in the field frame $\tilde{K}$ it is sufficient to perform a Lorentz transformation. The corresponding relations have the form\footnote{Note that the form of this transformation is similar to the \eqref{KKK}.}
\be\n{fff}
\begin{split}
&f^0=\gamma (\CAL{F}^{0}+\vec{V}\cdot \vec{{\CAL{F}}}) \, ,\\
&\vec{f} =\vec{{\CAL{F}}}+\gamma \CAL{F}^{0} \vec{V}+(\gamma-1)\dfrac{ \vec{V}\cdot\vec{{\CAL{F}}}}{V^2}\vec{V}\, .\,
\end{split}
\ee
Here $\gamma=(1-V^2)^{-1/2}$.

We use expressions \eqref{FFT}, \eqref{FFS} and \eqref{FFN} for the force $\ts{\CAL{F}}$ in the instantly comoving frame $K$ to obtain the expressions for the force $\ts{f}$  for the $T$-, $S$- and $N$-fields.

\subsubsection{$T$-field}

 In the case of the $T$-field, we find that

 \be
 f^{0}_{T} = \beta\gamma Mr_{+}\hh\vec{f}_{T} = \frac{1}{3}\beta\gamma^{2}\,\vec{J}\times\vec{V}\,.
 \ee
The torque for this case is
\be
\vec{\mathcal{T}}_{T} = \frac{1}{6}\beta\gamma^{2}\bigg[
M \vec{V}\times(\vec{V}\times\vec{J})
+\dfrac{8}{5M} (\vec{V}\cdot\vec{J})
(\vec{V}\times\vec{J})\bigg]\,
\ee

\subsubsection{$S$-field}

To calculate ${f}^{\mu}$, the following relations are useful
\be 
\begin{split}
& \vec{V}\cdot \vec{U}=\gamma (\vec{V}\cdot \vec{N})\, ,\\
&\vec{V}\cdot \vec{\CAL{F}}=-\beta \gamma(\vec{V}\cdot\vec{N})
\big[ \gamma Mr_{+}(\vec{V}\cdot \vec{N})
+\dfrac{1}{3}\, \vec{J}\cdot (\vec{V}\times\vec{N}) \big] \, .
\end{split}
\ee
One gets

\ba \n{fs}
&f^{0}_{S}=-\dfrac{1}{3}\beta\gamma^{2}(\vec{V}\cdot\vec{N})\, \vec{J}\cdot\big(\vec{V}\times\vec{N}\big)\, ,\\
&\vec{f}_{S}=-\beta\gamma 
(\vec{V}\cdot\vec{N})\Bigg[Mr_{+}
\vec{N}-\dfrac{1}{3}
\Big(
\vec{J}\times\vec{N} \\
&+\dfrac{\gamma-1}{V^2}
\Big[
(\vec{V}\cdot\vec{N})(\vec{J}\times\vec{V})+
(\vec{J}\cdot(\vec{N}\times\vec{V}))\vec{V}
\Big]
\Big)
\Bigg]
\, ,\\
\ea

\subsubsection{$N$-field}

The force acting on a black hole immersed in the $N$-field can be written as follows
\be
f^{\mu}_{N}= (1+\vec{N}\cdot\vec{V})\Big(f^{\mu}_{T} + \frac{f^{\mu}_{S}}{\vec{N}\cdot\vec{V}}\Big)\,,
\ee
where $f^{\mu}_{T}$ and $f^{\mu}_{S}$ are defined above.

\section{Special cases}

Let us now consider some special cases of motion of the rotating black hole in a homogeneous scalar field.

\subsection{Black hole motion in $T$-field}

This case was discussed in \cite{Frolov:2023gsl}. Here we add only a few remarks.
Let us denote 
\be 
\vec{P}=m\gamma \vec{V}\, .
\ee
Then equations \eqref{MJ} and \eqref{fff} imply

\be \n{Teqns}
\begin{split}
&\dfrac{dM}{d\tau}=\beta\gamma^{2}Mr_{+} \, ,\\
&\dfrac{d\vec{J}}{d\tau}=\dfrac{\beta}{6M}\Bigg[
\vec{P}\times(\vec{P}\times\vec{J})+\dfrac{8}{5M^2}\Big((\vec{P}\cdot\vec{J})(\vec{P}\times\vec{J})
\Big)\Bigg]\, ,\\
&\dfrac{d\vec{P}}{d\tau}=-\dfrac{\beta}{3M}\gamma(\vec{P}\times\vec{J})
\end{split}
\ee
These equations allow one to show that

\be \n{SOE}
\begin{split}
&\vec{P}^2=\mbox{const.}\,\hh (\vec{J}\cdot\vec{P})=\mbox{const.}\, ,\\
& \dfrac{1}{2}\dfrac{d\vec{J}^2}{d\tau}=-\dfrac{\beta}{6M}\Big[ 
\vec{P}^2 \vec{J}^2 -(\vec{J}\cdot\vec{P})^2
\Big]\, .
\end{split}
\ee 
The last relation also directly follows from \eqref{DDJJ}.

Let us note that $\gamma\ge 1$ and $r_+\ge M$. Then, the first equation in \eqref{Teqns} implies that
\be \n{ineq}
\dfrac{1}{M^2}\dfrac{dM}{d\tau}\ge\beta \, .
\ee 
Let $M=M_0$ be the initial mass of the black hole at $\tau=0$. Denote 
\be 
\tau_1=\dfrac{1}{\beta M_0}\, .
\ee 
Then by integrating the inequality \eqref{ineq}, one arrives to the conclusion that the mass $M$ becomes infinity at some time $\tau_0\le \tau_1$. This is a generic property of the behavior of a black hole in the $T$-field, which singles this case out and distinguishes it from the cases of the $S$- and $N$-fields. Special solutions of the equations of motion of the black hole in $T$-field can be found in \cite{Frolov:2023gsl}.

\subsection{Black hole motion in $S$-field}

\subsubsection{Black hole motion  transverse to the field} 

For starters, we assume that the black hole moves in the direction orthogonal to the $S$-field. Since $(\vec{V}\cdot\vec{N})=0$, one has $f_S^0=0$ and $\vec{f}_S=0$. This means that the both the mass $M$ of the black hole and its velocity $\vec{V}$ are constant. The expression for the torque \eqref{TOR} takes the form
\be 
\vec{\CAL{T}}=\dfrac{1}{6}\beta \Bigg[
M\vec{N}\times(\vec{N}\times\vec{J})+\dfrac{8}{5M} (\vec{N}\cdot\vec{J}) (\vec{N}\times\vec{J})
\Bigg]
\ee
It is easy to see that
\be \n{ORT}
\vec{\CAL{T}}\cdot \vec{N}=0\, .
\ee 
Let us write the spin as follows
\be 
\vec{J}=J_{\parallel}\vec{N}+\vec{J}_{\perp}\, .
\ee 
Equation \eqref{ORT} implies that $J_{\parallel}$ is constant and one obtains the following equation
\be\n{JPER} 
\dfrac{d\vec{J}_{\perp}}{d\tau}=\dfrac{1}{6}\beta\Big[
-M \vec{J}_{\perp}+\dfrac{8}{5} \dfrac{J_{\parallel}}{M} (\vec{N}\times \vec{J}_{\perp})
\Big]\, .
\ee 
To solve this equation, we introduce coordinates $(y_1,y_2,y_3)$ in which
\be 
\vec{N}=(0,0,1)\hh \vec{J}_{\perp}=(J_1,J_2,0)\, .
\ee 
We denote
\ba
&K=J_1 +i J_2\, ,\\
&\Gamma=\dfrac{1}{6}\beta\Big(
M-\dfrac{8}{5}i \dfrac{J_{\parallel}}{M}
\Big)\, ,
\ea
and so equation \eqref{JPER} takes the form
\be 
\dfrac{dK}{d\tau}=-\Gamma K\, .
\ee
A solution of this equation is
\be \n{KKK1}
K=K_0 \exp(-\Gamma \tau)\, .
\ee 
The constant $K_0$ is defined by the initial condition $K|_{\tau=0}=K_0$. 
The obtained solution shows that the spin of the black hole 
precesses around the direction of the gradient of the field with frequency $\omega=|\Im(\Gamma)|$, while its magnitude exponentially decreases. The characteristic time of this process is $\sim 1/\Re(\Gamma)$.

\subsubsection{Black hole motion  parallel to the field} 

For motion parallel to the gradient of the field, one has
\be 
\vec{V}=V\vec{N}\hhh U_{T}=\gamma V\hhh \vec{U}=\gamma \vec{N}\, .
\ee
The 4D force is
\be 
f_S^0=0\hhh 
\vec{f}=-\beta \gamma V \Big[Mr_{+}\vec{N}-\dfrac{1}{3}\gamma (\vec{J}\times\vec{N})
\Big]\, .
\ee 
If the spin vector $\vec{n}$ is not parallel to the field, there exists a component of the force in the direction orthogonal to the field. For the motion along the field this component should vanish. For this reason we impose the condition that $\vec{n}$ is parallel to $\vec{N}$. Since both vectors have a unit norm, one has $\vec{n}=\vec{N}$. It is easy to see that the torque for this case vanishes and it is sufficient to impose this condition only at the initial time.
Equation \eqref{DDJJ} also shows that the spin $J$ is constant.

\begin{figure}[h!bt]
    \centering
      \includegraphics[width=0.45\textwidth, trim=0 0 14 0, clip]{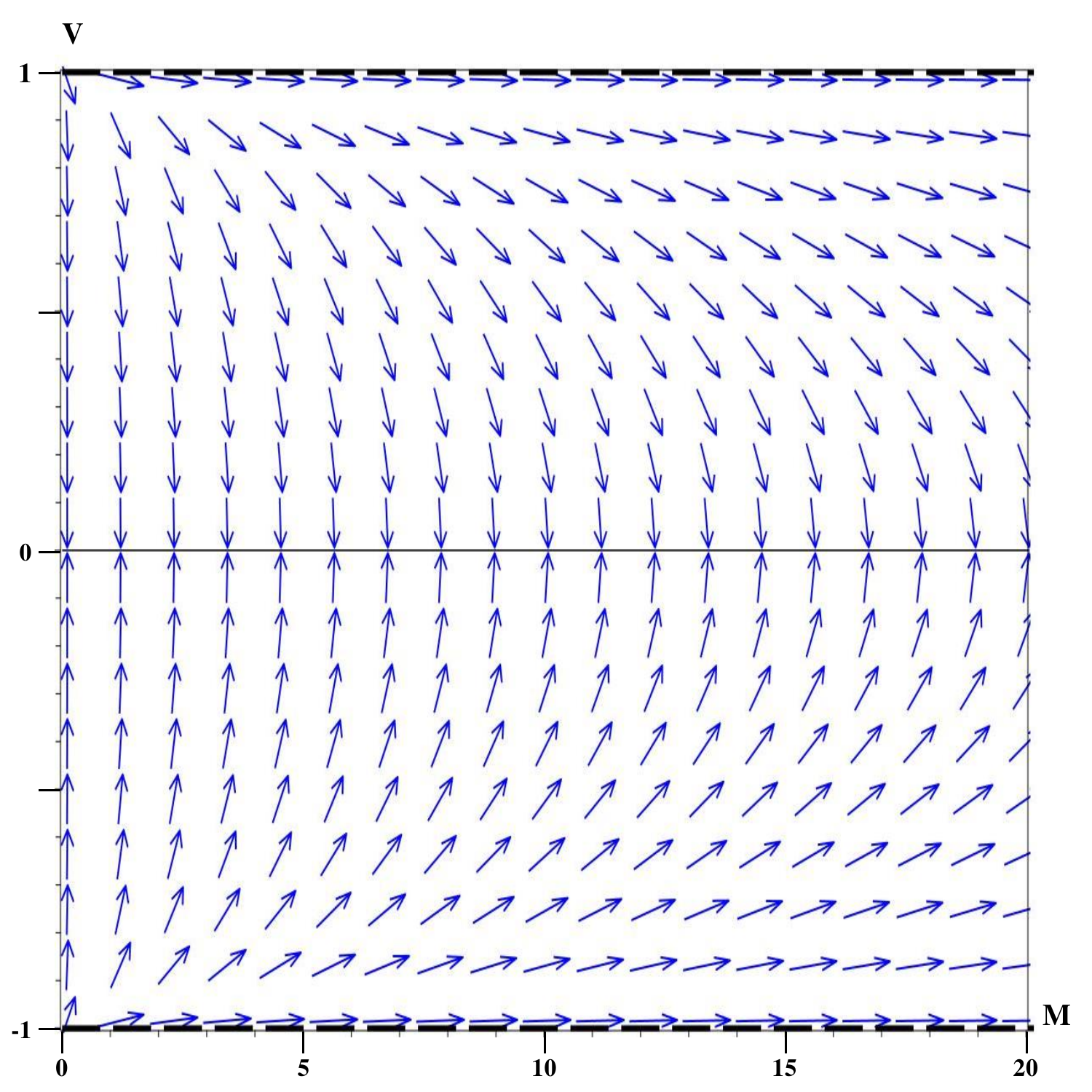}
    \caption{\n{MV_S} Phase space diagram of $M$ vs. $V$ for the $S$-solution with $\vec{V}=V\vec{N}=V\vec{n}$.}
\end{figure}

Equations \eqref{MJ} and \eqref{fs} imply
\be \n{MM}
\begin{split}
&\dfrac{d M}{d\tau}=\beta\gamma^2 V^2 Mr_{+}\, ,\\
&\dfrac{d (M\gamma V)}{d\tau}=-\beta\gamma V Mr_{+}\, ,
\end{split}
\ee
while condition $f_S^0=0$ means that
\be \n{M0}
\gamma M=\CAL{E}_0=\mbox{const.}\, .
\ee
The latter relation shows that the energy $\CAL{E}_0$ of the black hole is an integral of motion, and one has
\be \n{MfromEV}
M=\CAL{E}_0 \sqrt{1-V^2}\, .
\ee
Using this relation and \eqref{MM}, one obtains
\be \n{dMdV}
\begin{split}
&\frac{dM}{dV}=-\frac{MV}{1-V^2}\, ,\\
&\frac{dV}{d\tau}=-\beta V r_+\, .
\end{split}
\ee
A plot presented in Fig.~\ref{MV_S} shows the $(M,V)$ phase space associated with these equations.

This plot shows that for any initial mass $M$ and velocity $V$, the $S$-field produces a friction force which reduces the velocity of the black hole to zero. In this asymptotic limit, the growing mass $M$ reaches a finite final value equal to $\CAL{E}_0$.

Using \eqref{MfromEV} and the second equation in \eqref{dMdV}, one obtains the following equation for the black hole's velocity
\be \n{Zeqn}
\dfrac{dV}{d\tau}=-\beta \CAL{E}_{0}V \Big[ \sqrt{1-V^2}+
\sqrt{1-V^2-\dfrac{j^2}{1-V^2}}
\Big]\, .
\ee 
In the regime when the velocity of the black hole becomes small, one can approximate this equation and write
\be 
\dfrac{dV}{d\tau}=-\lambda V\hh 
\lambda=\beta\CAL{E}_0 \Big( 1+\sqrt{1-j^2}\Big)\, .
\ee
This equation shows that at later times, the velocity $V$ changes as
\be 
V=V_0 e^{-\lambda \tau}\, ,
\ee
and therefore it takes infinite proper time $\tau$ to reach the asymptotic value $V=0$.

\vspace{1cm}

\subsection{Black hole motion in $N$-field}

We consider the simplest case when the black hole moves within $N$-field in the direction parallel to it.
For this case, 
\be \n{VVNN}
\vec{V}=V\vec{N}\, .
\ee 
We also impose the condition
\be 
\vec{n}\times \vec{N}=0\, ,
\ee
which guarantees that the force component orthogonal to the field vanishes. For this case, if the relation \eqref{VVNN} is valid at an initial moment of time, it will always remains valid.

Under these assumptions, one has
\be 
\begin{split}
&U_T=\alpha\hhh \vec{U}=\alpha \vec{N}\hhh \CAL{F}^0=\beta \alpha^2 Mr_{+}
\, ,\\
&f^0=\beta \alpha Mr_{+}\hhh \vec{f}=-\beta \alpha Mr_{+} \vec{N}\, .
\end{split}
\ee 
Here 
\be 
\alpha=\sqrt{\dfrac{1+V}{1-V}}\, .
\ee
Also we note that the black hole under these conditions has constant spin, \ie $\vec{\mathcal{T}} = 0$.

We start with the following three equations

\ba
&\frac{dM}{d\tau}=\beta\alpha^{2} Mr_{+}\,,\\
&\frac{d(\gamma M)}{d\tau}=\beta\alpha Mr_{+}\,,\\
&\frac{d(\gamma MV)}{d\tau}=-\beta\alpha Mr_{+}\,.
\ea
It is easy to check that only two of these equations are independent, and that the third one follows from the first two.

\begin{figure}[h!bt]
    \centering
      \includegraphics[width=0.45\textwidth, trim=0 0 14 0, clip]{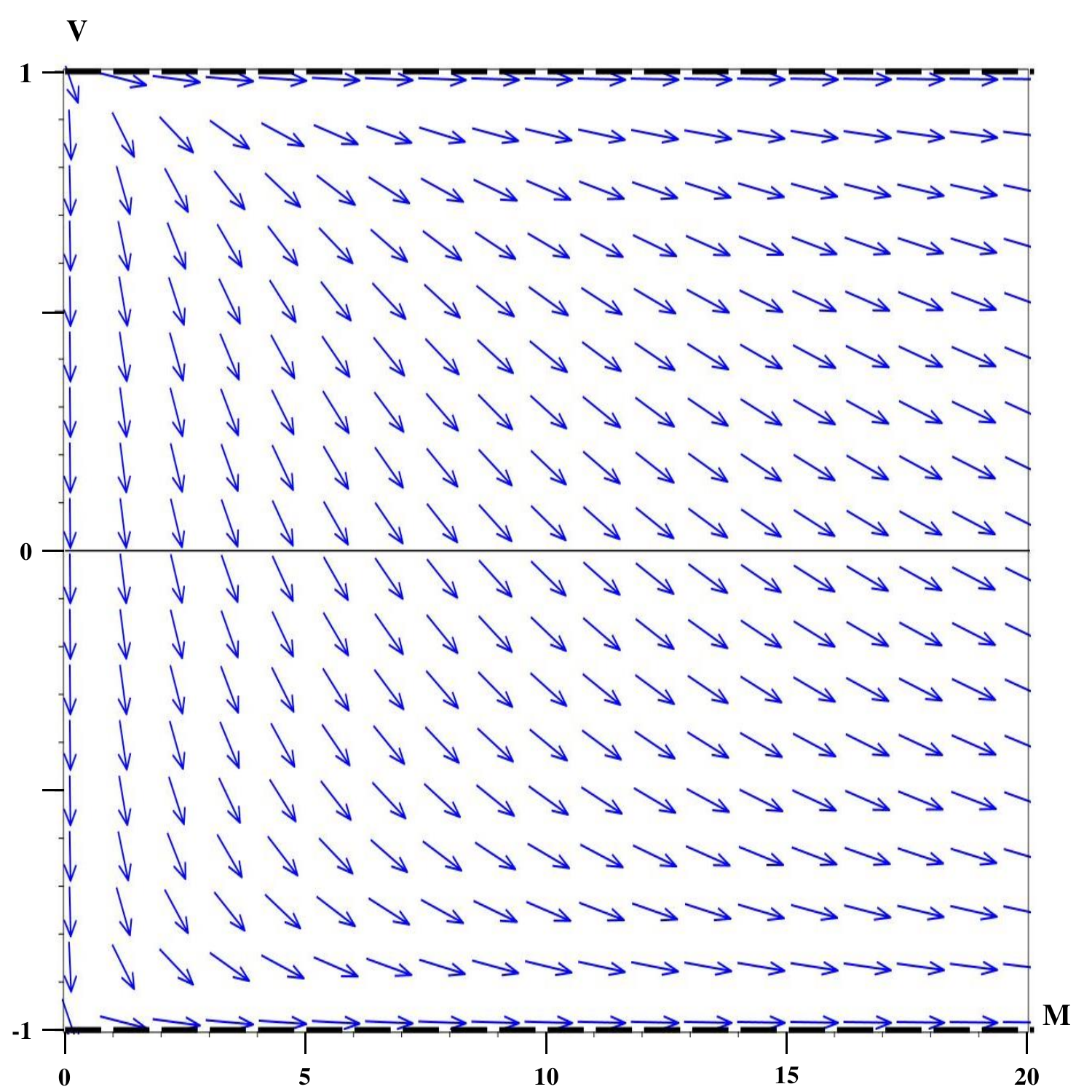}
    \caption{\n{MV_N} Phase space diagram of $M$ vs. $V$ for the $N$-solution with $\vec{V}=V\vec{N}=V\vec{n}$.}
\end{figure}

After some algebra, we arrive at the following system of differential equations for $M$ and $V$
\ba \n{MVN}
\frac{dV}{d\tau}=&-\beta r_{+}(1+V)^{2}\,,\\[8pt]
\frac{dM}{d\tau}=&\beta Mr_{+}\frac{1+V}{1-V}\,.
\ea
Combining them gives
\be \n{dMdVN}
\frac{dM}{dV}=-\frac{M}{1-V^{2}}\,.
\ee
The $(M,V)$ phase space for this equation is shown in Fig. \ref{MV_N}.
This plot shows that $V=V(M)$ is a monotonically decreasing function, which takes the value $V=1$ for $M=0$, and asymptotically reaches the value $V=-1$ when $M\to \infty$. One can also solve this equation to obtain the following relation between $M$ and $V$
\be \n{Malpha}
M=M_0\, \sqrt{\dfrac{1-V}{1+V}}\,.
\ee
Here $M_0$ is the mass of the black hole when it has zero velocity.

Using \eqref{MVN} and \eqref{Malpha} one can also obtain obtain a following equation for the change of the mass
\be \n{MMTT}
\dfrac{dM}{d\tau}=\beta M_0^2 \, \dfrac{r_+}{M}\, .
\ee 
Since $M < r_+\le 2M$ this relation shows that 
\be \n{INEQ}
\beta M_0^2 < \dfrac{dM}{d\tau}\le 2\beta M_0^2\, .
\ee

\section{Summary and discussion}

In this paper, we discussed effects connected with the motion of a rotating black hole in a static homogeneous scalar field of general configuration. It extends the results of the earlier publication \cite{Frolov:2023gsl} to the cases in which the vector of the field gradient is not only timelike, but can also be spacelike or null as well. Another difference from the previous publication is that a complete set of equations is obtained describing not only the change in magnitude of the black hole's spin, but also its orientation. 

We demonstrate that the mass of a rotating black hole moving in a scalar field cannot decrease, while the magnitude of its spin can never increase. As a result, the surface area (and hence the entropy) of the black hole never decreases as well. This result is in complete agreement with the second law of black hole physics. Although these results are general and valid for any $T$-, $S$- and $N$-field configurations, there exist very important differences between these cases. In the $T$- and $N$-fields, the mass of the black hole always grows and cannot remain constant. 

For motion within the $S$-field, the mass of the black hole can remain constant under the right conditions. This happens when either the black hole is at rest with respect to the field frame, or when it moves with velocity orthogonal to the direction of the field gradient. In the case where the black hole's velocity is parallel to the field gradient, the growth of the black hole's mass is accompanied by a decrease in its velocity. The final value of the black hole mass is finite and it is determined mainly by its initial energy. It takes infinite time to reach this asymptotic state in which $V=0$.

For the black hole moving in the $N$-field, its mass and the absolute value of the velocity $|\vec{V}|$ monotonically grow with time. At late time the velocity $V$ is negative, that is the black hole moves in the opposite to $X$ direction. In the limit $\tau\to\infty$ the velocity $V\to -1$. These results have a simple and natural explanation. The stress-energy tensor of the scalar field in the $N$-state has the form of a left-moving null fluid. The black hole absorbs its energy and momentum, and this results in the growth of it mass and velocity $|\vec{V}|$.

The case of the $T$-field is singled out by the following property: For a black hole with an arbitrary initial velocity (including the case when it is at rest), the mass of the black hole infinitely grows and (at least formally) it reaches an infinite value within a finite interval of time. This property was already discussed in the paper \cite{Frolov:2023gsl}. 
A simple explanation of this phenomenon is given as follows: The rate of change of the black hole's mass is proportional to its cross-section $\sim M^2$,  
\be 
dM/d\tau \sim M^2\, .
\ee 
The integral
\be 
\tau_f\sim \int^{M_f} \dfrac{dM}{M^2}\, ,
\ee 
which shows that the time $\tau_f$ when the mass reaches the value $M_f$  is finite for $M_f\to \infty$. This means that it takes the black hole finite time $\tau$ to reach an infinite value of mass.
Let us note that for $N$-field, the situation is qualitatively different. 
Equation \eqref{INEQ} shows that at a late time, the rate of the change of the black hole mass is practically constant. The explanation of the difference with the $T$-field case is that for the $N$-field, when $V\to -1$, its null fluid energy flux as observed in the black hole frame is highly suppressed  by a Doppler shift  (see \eqref{TRED}). This effect exactly compensates the effect of the black hole's cross-sectional growth.

In many aspects, the motion of the black hole in the $S$-field is similar to the motion of a rotating black hole in a homogeneous electromagnetic field discussed in \cite{Frolov:2024xyo}, with one important difference: In the latter case the mass of the black hole never changes. 

In our discussion, we considered the simplest version of a scalar field. Namely, we assumed that it is a minimally-coupled massless field obeying a linear  equation. It would be interesting to use the developed approach and apply it to study of the interaction of rotating moving black holes with more complicated fields, such as ghost condensate, khronon and aether fields.

\bibliography{KERR}

\end{document}